# You Only Live Twice or "The Years We Wasted Caring about Shoulder-Surfing"

Joseph Maguire
School of Computing Science
University of Glasgow
Glasgow, G12 8QQ
Scotland, United Kingdom
www.dcs.gla.ac.uk/∼joseph
*joseph@dcs.gla.ac.uk*

Karen Renaud
School of Computing Science
University of Glasgow
Glasgow, G12 8QQ
Scotland, United Kingdom
www.dcs.gla.ac.uk/∼karen
*karen@dcs.gla.ac.uk*

**Passwords are a good idea, in theory. They have the potential to act as a fairly strong gateway. In practice though, passwords are plagued with problems. They are (1) easily shared, (2) trivial to observe and (3) maddeningly elusive when forgotten. While alternatives to passwords have been proposed, none, as yet, have been adopted widely. There seems to be a reluctance to switch from tried and tested passwords to novel alternatives, even if the most glaring flaws of passwords can be mitigated. One argument is that there is not enough investigation into the feasibility of many password alternatives. Graphical authentication mechanisms are a case in point. Therefore, in this paper, we detail the design of two prototype applications that utilise graphical authentication mechanisms. However, when forced to consider the design of such prototypes, we find that pertinent password problems eg. observation of entry, are just that: *password* problems. We conclude that effective, alternative authentication mechanisms should target authentication scenarios rather than the well-known problems of passwords. This is the only route to wide-spread adoption of alternatives.**

*authentication, shoulder-surfing, economics*

## 1. INTRODUCTION

The problems surrounding passwords are not new: they are the stuff of legend, as evident from the following tale:

*Two brothers lived in a small town. Cassim was wealthy and wanted for nothing, Ali Baba was poor and cut wood tirelessly to support his family. While cutting wood in the forest one day, Ali Baba spotted a flock of fierce men on horses heading his way and so hid in a nearby tree. The leader walked right-up to the tree and uttered a simple, secret word. A huge rock rolled-back to reveal a vast cave, covered in treasure.*

*Stunned and shaken, Ali Baba stayed silent in the tree as the men entered the cave and the rock rolled shut. The men eventually left, at which point Ali Baba leaped from the tree and spoke the simple, secret word. The rock rolled-back and Ali Baba grabbed as much treasure as he could and ran back to his family.*

*Cassim, shocked by Ali Baba's sudden wealth, demanded answers. Ali Baba recounted the story and agreed with his brother to return in the morning. Cassim, always the greedier, set out earlier. Once* *at the cave, Cassim spoke the simple, secret word and, to his amazement, the rock rolled-back. Cassim entered and the rock rolled shut, he began grabbing as much treasure as possible. However, when Cassim wanted to leave, he could not recall the simple, secret word. He tried in vain to remember, spouting word after word.*

*The flock of thieves returned to the cave to deposit more treasure. The leader spoke the simple, secret word. The rock rolled-back to reveal Cassim cowering in the corner of the cave. Cassim never left.*

The tale of Ali Baba and Cassim is more commonly known as *Ali Baba and the Forty Thieves* from *1001 Arabian Nights*.

Some lessons emerge from this narrative:

- The first paragraph illustrates that passwords are *not resistant to surveillance*. Ali Baba, after all, was able to enter the cave because he was eavesdropping. The problem still persists today, as attackers can easily observe individuals entering their passwords by shoulder-surfing.

Passwords need to be resistant to eavesdropping and observation, otherwise their strength is weakened. A password is an **authentication secret**, it needs to be kept private and safe.

- The second paragraph showcases the authentication secret being *shared with others*. Ali Baba is able to communicate the password to his brother and, as a result, Cassim was also able to access the cave. Passwords should be known only to the individual accessing the cave, no-one else. Otherwise, it weakens the strength of the password as it is no longer a secret.

- The third paragraph demonstrates that passwords are *not memorable*. Cassim's life literally depended on his remembering the password that Ali Baba shared with him. A good password is a lengthy string of gibberish with no meaning. The lack of meaning is what makes such passwords difficult to remember. If a user does not write it down, they will almost certainly reuse it. While a memorable password has meaning, it also makes it predictable as it has some structure or pattern. Making it easily compromised through brute-force or social-engineering attacks.

- The final paragraph exemplifies the fact that users and organisations only know when the attacker is unsuccessful. When the leader of the thieves rolls back the rock, he is confronted with Cassim. The authentication secret has been compromised, the password is pointless. In modern-day electronic systems there is often no way to detect that an attacker has successfully used another person's password.

These problems have always plagued passwords, even when they were first deployed on the Compatible Time Sharing System (CTSS) in 1962 (Corbató *et al.*, 1962). The key difference between then and now is that CTSS was aimed at scientists and engineers. The password pitfalls could be combatted with training and education. Passwords are undeniably powerful, if used properly. Organisations can craft secure locations, make surveillance difficult, and use security policies and training to mitigate the aforementioned problems. The real problem is everyday users, managing an increasing number of passwords to achieve more and more tasks. Popular products like the Nintendo Wii and Apple iPhone have pushed passwords into untrained hands and spaces, never initially envisioned or planned for.

What is needed is a *password for plebs*. Such an authentication approach would be tailored to the needs of the average user of popular consumer products such as the Nintendo Wii and Apple iPhone. Such an approach should tackle the problems of *observation*, *memorability* and *sharing of authentication secrets*. This authentication approach would be targeted at consumers and popular products rather than professionals and tele-type terminals. The approach we proposed was a *graphical authentication mechanism* called Tetrad (Renaud & Maguire, 2009). This mechanism utilised an image-based authentication secret that users could enter in a way that was resilient to surveillance. Tetrad appeared to solve a number of the problems of passwords: the images were memorable, observation of entry had the potential to not leak the authentication secret and it is cumbersome to note-down a picture password. We thought we had found a viable alternative.

The initial prototype we constructed of Tetrad showed promise. Unfortunately, Tetrad, like many other graphical authentication mechanisms, has yet to be adopted by any commercial product. Plebs still use passwords and those passwords still suffer from the age-old problems. For example, the Nintendo Wii and Apple iPhone have yet to make any changes to their authentication approach. In particular, neither have bothered to tackle the problem of shoulder-surfing even though both rely on an on-screen keyboard for password entry, easily observable if users enter their passwords in full view of onlookers. This makes observation trivial and the authentication secret public. Is this evidence of negligence or wisdom? Here we will explain that it is actually a perfectly justified decision.

Dunphy *et al.* argues the reason for slow uptake in graphical authentication approaches, like Tetrad, is due to designers and researchers not tackling practical issues such as deployment (2010). This is certainly true of the initial incarnation of Tetrad. The initial implementation comprised only the mechanism itself; there was no registration process and it served no authentication purpose. Since it was not protecting any application it could be argued that the evaluation we carried out to prove its efficacy was unrealistic. The implementation did not target an actual platform but rather a mythical Apple television context of use. Furthermore, the mechanism relied on a stock collection of celebrity faces. Personal images were avoided because initialising or 'bootstrapping' a graphical authentication with them is hard. Personal images need to be collected from the user as well as prepared and processed. The use of celebrity faces simplified the situation, as they are readily available, but such images suffer from predictability issues, so the choice was less than optimal. These prickly deployment issues were neatly avoided in our early prototype.

In this paper we discuss the prototyping of two versions of Tetrad. We will refer to them as *Jack* and *Jill*. Both versions rely on personal images from Facebook. We will discuss deployment choices related to image bootstrapping procedure and the choice of application both mechanisms were bundled with. We will outline the authentication scenario and explain how the mechanism interacted with the application.

The following section presents related research, regarding the problems of passwords and the concerns of observability of graphical authentication secrets. The next section outlines the design of the application, followed by a section detailing the image-bootstrapping procedure for both versions of Tetrad. Finally, we discuss the problems we encountered in image-bootstrapping and address the relevancy of obfuscation in graphical authentication. The paper concludes that shoulder-surfing may well be a non-issue in authentication design.

## 2. RELATED RESEARCH

There are many problems with passwords. Good passwords are system-generated lengthy strings of gibberish with no meaning. However, such passwords are difficult for a user to manage in the short-term, never-mind store long-term. Users, therefore, adopt coping mechanisms (Chiasson *et al.*, 2009), such as writing passwords down or reusing them, to avoid inconvenience. User-generated passwords are an alternative option but people are 'lazy' and often create short and memorable authentication secrets (Riddle *et al.*, 1989). Memorable passwords have meaning, i.e. pattern and structure, that brute-force and social-engineering attacks can easily exploit. One solution to all these problems is an alternative authentication approach, e.g. graphical passwords.

Graphical passwords are based on images rather than characters. Humans use a dual-coding approach to process verbal and visual stimuli, generating analogue codes for visual information and symbolic codes for verbal (Paivio, 1990). This approach results in images having superior recognition for the words that name them (Nelson *et al.*, 1976). Individuals are also good at recognised previously encountered images, even in pairs containing a distractor (Shepard, 1967).

A concern with graphical passwords is the predicability of passwords created by users. Users are influenced by elements such as attraction, race and gender when selecting faces. Therefore, an attacker can potentially identify likely images that form part of a user's authentication secret by simply knowing something about the user. This makes sourcing images or bootstrapping a mechanism with images, difficult, as some collection of images may not be suitable. Celebrity images for example, a male user may select attractive female users around his age. The pool of images would be nearly the same for every similar user, making it easy for an attacker to identify a likely password. Graphical passwords based on personal images have strong memorability (Tullis & Tedesco, 2005) and may be alternative solution, as they would be tailored to the user. Furthermore, social network services have made accessing and using such personal images much easier. Therefore, we decided to construct Tetrad prototypes that used images of faces, extracted from personal images, sourced from Facebook.

## 3. APPLICATION DESIGN

The first step in the design of the application was to define an **authentication scenario**. The envisioned scenario for the prototype was a digital content store. Purchasing a movie in front of friends and family is a familiar authentication scenario for many. Users peruse the catalogue, pick a product and then enter a password to purchase it. Current consumer products expect users to enter their password using an on-screen keyboard, essentially exposing it to onlookers. This burdens the user with the decision of either sacrificing their authentication secret to maintain good vibrations or being the paranoid android that demands everyone exits the room.

Tetrad is designed to tackle this scenario, allowing an individual to authenticate without worrying about onlookers. The mechanism presents 45 images within a grid and asks users to position four secret images so that they align either horizontally, vertically or diagonally. Interaction does not involve moving individual images but manipulating rows and columns of them, resulting in entry being resilient to observation from onlookers.

While numerous scenarios were considered, including a password manager and location tracker, the conclusion was to continue with the original one, a digital content store. A digital content store is a natural fit for Tetrad as both heavily depend on screens. Digital stores rely on high-quality screens that are spacious enough for a user to peruse a product catalogue and splendid enough to showcase content. Tetrad, equally, relies on screens with enough space and quality to showcase images.

The alternative scenarios were seriously considered but ultimately rejected. Deploying a password manager or location tracker, for example, puts the user at risk. The application will, inevitably, contain bugs and flaws. When it fails, and it will, the user's passwords and locations would be exposed, an unacceptable risk. A digital store puts content at risk, not the user, a preferable scenario for evaluation.

Therefore, the resulting application was a *digital content store and accompanying media player*, designed for the iPhone. Tetrad was implemented in Objective-C and originally only targeted Mac OS X. The implementation made extensive use of key frameworks, such as Core Animation, and users interacted with it using an Apple Remote Control. The prototype was designed for use on a television and was aimed at tackling the aforementioned authentication scenario.

Translating the implementation to the iPhone required additional effort but was aided by the fact that the iPhone's operating system is essentially Mac OS X. Interaction had to be re-thought as the remote control was no longer relevant. Remote control events thus became multi-touch gestures. Users swipe up, down, left and right on a particular row or column to reposition images. A double-tap gesture submits an authentication attempt.

The second step in the design process was to define the **consequences of authentication**. Users need to understand the actions that follow authentication. Screen real-estate is a premium on smartphones, making text descriptions and instructions a luxury. Consistent interaction should ensure that the mere presentation of the authentication mechanism communicates what actions follow. Sky serve as a bad example, using a single PIN for (1) accessing age-rated content and for (2) purchasing premium content. Sky would argue the purpose is the same in both cases, confirming the user is the bill-payer. However, PIN entry actually has two possible consequences, leading to ambiguity and confusion. Users may accidentally purchase premium content while assuming they are accessing age-rated subscription content.

The consequences of authentication in the application had to be clear. Tetrad was coupled with a content store, consequences were therefore limited to either (1) payment or (2) access. The first approach is the one adopted by iTunes while the second is akin to Spotify and Netflix.

The latter approach was favoured, as the application comprised of a digital store and media-player. Purchased content would be kept within the application and individuals would use the application's media-player to consume content. Tetrad regulated access to application usage.

This design was used for both prototypes. What separated the prototypes was the image-bootstrapping procedure used.

## 4. IMAGE BOOTSTRAPPING

Personal images were sourced from Facebook. The first step in both prototypes required users to connect their Facebook account with Tetrad and grant permission for the mechanism to access personal images of friends.

### 4.1. Jack — Face Recognition

The first challenge for the Jack prototype was to analyse the images. The iPhone operating system, at the time, did not provide support for detecting faces within images. The Open Source Computer Vision Library or OpenCV was used to process personal images on the iPhone. The cross-platform library has a vibrant community, providing documentation and support for any number of generic image processing algorithms, including detecting faces within images.

The library was bundled with the application. Once the user had connected Tetrad with Facebook, the mechanism submitted a query to the social network for the user's friend list. The response to the query was used to download the profile picture of all the user's friends. The image processing procedure was multi-threaded in an attempt to reduce the time taken to process the images. The application spawned multiple threads at once. An individual thread, dealt with a single friend. The thread downloaded a friend's profile picture, analysed it for faces, returned an array of possible faces and extracted a single face from the image and stored it on the device. The process did not recognise faces, i.e. the face detected in the profile picture, is not necessarily of the profile owner.

The process simply selected the first element in the array and used the coordinate information to locate the face within the image and extract it. The thread completed by adding an entry to list, the entry included the friend's name and the face extracted. The instructions requested users to select 45 friends. The table-view remained responsive while image-processing was happening. Therefore, users did not need to wait until all personal images were processed, they could select 45 images quickly and move-on.

The next-step requested users to select 4 of the 45 images, as their authentication secret. The user simply scrolled through a table-view of the 45 images and tapped the entries they wanted to make their authentication secret. The final step was to order the sequence of the 4 images. Users were instructed to memorise the extracted images and sequences.

Once the user was content with the sequence, they could complete registration.

### 4.2. Jill — Tags

The second prototype took a different direction to image-bootstrapping, favouring a less computational expensive approach. Instead of downloading and analysing images to detect faces, the mechanism relied on 'tags'. Tags are used to identify friends within images on Facebook. The social network service provides a tagging tool, affording users the ability to drag a box around a friend's face within an image. Users can then easily see all the photographs that include a specific friend.

The Jill prototype utilised these tags, rather than attempting to analyse and detect faces itself. The registration process was compressed into a single-stage. The aim was to make the registration process feel zippier and more responsive, than earlier versions.

The user's friend-list was presented in table view, with a table entry including the friend's name and profile picture. Users were instructed to select 45 friends, with 4 of them making-up their authentication secret. All a user had to do was simply tap a table-entry to select a friend. When a friend was selected, users had the option to add that friend to their authentication secret. They also had the option to order the sequence of their authentication secret, at the same time. When the user had selected 45 friends and made 4 of them their authentication secret - a button was enabled that allowed them to complete the registration process.

### 5. DISCUSSION

Designing and developing prototypes of Tetrad that relied on personal images only served to emphasise the difficulty in deploying the mechanism in the real world. The initial incarnation of the authentication approach may have showed promise but the viability of both prototypes is questionable.

The authentication approach relied on faces. The prototypes extracted these faces from personal images, sourced from Facebook. Putting aside the actual usability of either prototype and merely focusing on the practicality, neither seemed worthy of further investigation. Bootstrapping either prototype with images was incredibly expensive in terms of time and resources, especially so when contrasted with the tried and tested, as well as convenient, alphanumeric approach.

The prototypes extracted and used faces in different ways. The first prototype, titled 'Jack', actually detected faces within personal images and extracted them while the second prototype, titled 'Jill', used user-generated tags from Facebook to extract faces from personal images.

The image-bootstrapping procedure for Jack was incredibly expensive as images had to be analysed to extract faces. This consumed significant resources on the user's device and took some time. Alternatively, image processing of personal images could have occurred in the cloud instead of on the device. The cloud option is desirable as image processing and analysis is an intensive task. Amazon can provide powerful cloud computing instances that allow for multi-threaded image processing operations. The cloud computing instance could locate and extract faces within a few seconds, contrasted with several minutes on a mobile device. The extracted faces could be downloaded directly to the device and the instance could be discarded. The main concerns with the cloud computing approach is the bottleneck in downloading images, the cost associated with the operation and the privacy risk to the user. These concerns led us to favour the second option, processing the personal images on the user's device.

Meanwhile, the second prototype, Jill, avoided costly computation by using tags from Facebook to extract faces. Although the prototype avoids costly computation it still requires a complex registration process. The user needs to spend time selecting images for use within the mechanism and they have to wait for those images to download. Contrast this with alphanumeric authentication, where a user can quickly tap out a text password in a few seconds.

The biggest mistake security researchers can make is assuming that user effort were free (Herely, 2009). The benefit of shielding an authentication secret, for example, must not come at an exhorbitant cost. Therefore, the real question becomes: How big a threat is shoulder-surfing? The threat needs to be fairly big to be worth the hassle of dealing with Tetrad, instead of alphanumerics. The prototypes forced thought on deployment issues, such as how the mechanism would be bootstrapped with personal images and what sort of application it would protect.

The original scenario we envisioned was a user purchasing a movie in front of friends and family. The reality, however, is that most users do not care if these people know the password they use to purchase movies. Users want to share a movie in their living room, a far more valuable and intimate thing than a password. If a friend or family member *does* use someone's else password, then they will work it out. The same way they would, for example, if someone took the last beer or slice of pizza. Shoulder-surfing in this context is a non-issue, something we failed to recognise.

Therefore, for the prototypes we envisioned the scenario of an individual accessing a digital store, being exposed to the wider world. For example, Ppurchasing a movie on a smartphone while having a coffee at Starbucks or waiting for train in a station. These places are bursting with strangers and a user would not want any of them to observe entry of their password. However, users are likely to authenticate on a 4" inch screen, not a 40" one. Therefore, they can position themselves and the screen to prevent observation.

Lets say a stranger does, somehow, manage to observe the password being entered in such a space. What are the actual consequences, in the case of a digital store? If an attacker does manage to purchase several items from a service, such as iTunes, Apple can de-activate the purchases.

Therefore, maybe the focus on digital stores was the wrong scenario. Tetrad could have been used to protect email or a financial service. However, no individual would check their personal email account or personal finance account on a television in front of friends and family. Not necessarily because the information is sensitive but because it is boring and uninteresting. Large screens are for sharing and although some users might want to read email and check balances on a big-screen, they would wait until the television is free, rather than interrupting a movie.

Constructing prototypes that are meant to be the manifestation of a practical, deployable product require us to focus on the actual application used, the people using it and the places where it will be used. Authentication is a package, it is not something that is spooned on top of a finished product, it is part of the product. The time and energy taken to create a Tetrad secret is not necessary worth the hassle because shoulder-surfing is not really something to be concerned about in many contexts.

In the end we didn't actually tackle an authentication problem; we tackled a password problem. Tetrad was not designed thinking about the users and the tasks they wanted to complete but in order to right the wrongs of passwords. Tetrad was burdened with the legacy of the password approach. The design process did not start with: people want to buy movies in the living room. Instead it started with: these are the problems with passwords in the living room.

## 6. CONCLUSION

*You only live twice, or so it seems.*
*One life for yourself and one for your dreams.*

*You drift through the years and life seems tame*
*Till one dream appears and love is its name*

Sung by Nancy Sinatra, the lyrics above are from the John Barry classic, 'You only live twice'. They are inspired by the film and novel of the same name and the lyrics speak to the idea that an individual 'lives twice'. The life they led and the life they lead after facing death.

We thought we were testing Tetrad. The surprising result of the evaluation was that shielding authentication secrets from surveillance, in many contexts, is misguided at best and pointless at worst.

The route to wide-spread adoption of authentication alternatives is though solving an *authentication* scenario. Authentication is a package. The design process should start by considering every aspect of that package: people, places and purpose. Solving the problems of passwords doth not a successful authentication mechanism make.